\documentclass[aip,pof,reprint]{revtex4-1}
\usepackage[dvipsnames,rgb,dvips]{xcolor}
\usepackage{graphicx}
\usepackage{dcolumn}
\usepackage{float}
\usepackage{overpic}
\usepackage{amsmath,amssymb}
\usepackage{bm}

\usepackage{todonotes}
\usepackage{units}
\definecolor{figred}{RGB}{228,26,28}
\definecolor{figblue}{RGB}{55,126,184}
\definecolor{figgreen}{RGB}{77,175,74}
\definecolor{figpurple}{RGB}{152,78,163}

\newcommand{\ve}[1]{\ensuremath{\mbox{\boldmath$#1$}}}
\newcommand{\ma}[1]{\ensuremath{\mathbb{#1}}}
\newcommand{\ku}{\ensuremath{\mbox{Ku}}}

\newcommand{\rd}{\ensuremath{\mathrm{d}}}
\newcommand\transpose{^{\scriptstyle\mathrm T}}
\newcommand\nn{\nonumber}

\newcommand{\tauK}{\ensuremath{\tau_{\mbox{\tiny K}}}}
\newcommand{\ar}{\alpha}

\newcommand{\nrod}{\ensuremath{\ve{n}_\mathrm{rod}}}
\newcommand{\ndisk}{\ensuremath{\ve{n}_\mathrm{disk}}}

\def\clap#1{\hbox to 0pt{\hss#1\hss}}

\newcommand{\eqnlab}[1]{\label{eq:#1}}
\newcommand{\figlab}[1]{\label{fig:#1}}

\newcommand{\eqnref}[1]{(\ref{eq:#1})}

\newcommand{\Eqnref}[1]{Eq.~(\ref{eq:#1})}
\newcommand{\Figref}[1]{Fig.~\ref{fig:#1}}

\newcommand{\seclab}[1]{\label{sec:#1}}
\newcommand{\secref}[1]{\ref{sec:#1}}
\newcommand{\Secref}[1]{Section~\ref{sec:#1}}

\begin{document}
\title{Shape-dependence of particle rotation in isotropic turbulence }
\author{M. Byron}
\affiliation{Civil and Environmental Engineering, University of California, Berkeley, USA}
\affiliation{NORDITA, Roslagstullsbacken 23, 11421 Stockholm, Sweden}
\author{J. Einarsson}
\affiliation{Department of Physics, Gothenburg University, 41296 Gothenburg, Sweden}
\affiliation{NORDITA, Roslagstullsbacken 23, 11421 Stockholm, Sweden}
\author{K. Gustavsson}
\affiliation{Department of Physics, Gothenburg University, 41296 Gothenburg, Sweden}
\affiliation{NORDITA, Roslagstullsbacken 23, 11421 Stockholm, Sweden}
\author{G. Voth}
\affiliation{Department of Physics, Wesleyan University, Middletown, Connecticut 06459, USA}
\affiliation{NORDITA, Roslagstullsbacken 23, 11421 Stockholm, Sweden}
\author{B. Mehlig}
\affiliation{Department of Physics, Gothenburg University, 41296 Gothenburg, Sweden}
\affiliation{NORDITA, Roslagstullsbacken 23, 11421 Stockholm, Sweden}
\author{E. Variano}
\affiliation{Civil and Environmental Engineering, University of California, Berkeley, USA}
\affiliation{NORDITA, Roslagstullsbacken 23, 11421 Stockholm, Sweden}

\begin{abstract}
We consider the rotation of neutrally buoyant axisymmetric particles suspended in isotropic turbulence. Using laboratory experiments as well as numerical and analytical calculations, 
we explore how particle rotation depends upon particle shape.  
We find that shape strongly affects orientational trajectories, but that it has negligible effect on the variance of the particle angular velocity.  Previous work has shown that shape significantly affects the variance of the 
tumbling rate of axisymmetric particles. It follows that shape affects the spinning rate in a way that is, on average,  complementary to the shape-dependence of the tumbling rate.  We confirm this relationship using direct numerical simulations, showing how tumbling rate and spinning rate variances show complementary trends for rod-shaped and disk-shaped particles.  
We also consider a random but non-turbulent flow. This allows us to explore which of the features observed for rotation in turbulent flow are due to the effects of particle alignment in vortex tubes. 
\end{abstract}
% 05.40.-a Fluctuation phenomena, random processes, noise, and Brownian  motion
% 92.60.Mt Particles and aerosols
% 05.60.Cd Classical transport
% 45.50.Tn Collisions
% 47.27.-i Turbulent flows
% 05.40.Jc Brownian motion
% 47.55.Kf Particle-laden flows
% 47.27.eb Turbulence - Statistical theories and models
% 47.27.Gs Isotropic turbulence; homogeneous turbulence

\maketitle

\section{Introduction}
Non-spherical particles moving in turbulent flows are of fundamental importance for many different scientific problems. 
Examples are rain initiation by ice crystals in turbulent clouds \cite{Pru78}, fiber suspensions 
\cite{Lun11}, grain dynamics in accretion disks \cite{Wil08}, 
and pattern formation on the surface of turbulent and complex flows \cite{Wil09,Wil10a,Wil11}.
Our work is motivated by the problem of describing plankton dynamics in aquatic ecosystems. 
The dynamics of plankton in turbulent flows are of interest because plankton occupy the lowest marine trophic level and also significantly affect the contribution of the ocean to the global carbon budget \cite{sig00,jia10}.
These vital ecological functions are inextricably tied to the mechanical interactions between individual plankters and the complex flow environment they occupy. Rotation of planktonic organisms is critical for chemotaxis \cite{mac72,ish83}, and plays an important role in diffusive nutrient uptake \cite{ngu11}.  The physical and biological effects of rotation continue to be a subject of inquiry, along with other kinematic factors that have clearly been shown to influence feeding, reproduction, and predator-avoidance 
\cite{Kio08,Den95,Den10}.

Rotating axisymmetric particles may tumble, or spin, or both. The term tumbling denotes the orientational dynamics of the symmetry axis of the particle, while the term spinning denotes the rotation of the particle around its own symmetry axis. Tumbling of non-spherical particles has received attention recently \cite{Shi05,Par12,Che13,Gus14}, but spinning is 
an equally important part of rotation.       
For example, inertial particles do not simply sample fluid vorticity and strain, but rather extract angular momentum, transport it, and `return' it to the fluid phase. How this occurs likely depends upon how the total angular velocity of the particle is distributed between tumbling and spinning. 

The rotations of small particles suspended in linear shear flows have been studied intensively, both theoretically \cite{Jef22,Sub05,Sub06,Ein14,Lun10} 
and experimentally \cite{Ein13}.
In this context spinning is referred to as \lq log rolling\rq{}.
Here we consider particles suspended in turbulent flows and evaluate the effects of turbulence upon rotation, tumbling, and spinning. 

We consider simple shapes (spheroids and cylinders) with uniform distribution of mass, and particle sizes within either the dissipative or inertial subrange of turbulence. These ranges correspond roughly to the lower and upper bounds of body size of planktonic organisms.  Two representative groups of plankton, spanning this range of length scales, are diatoms 
(class Bacillariophyceae, $20$-$200$\, $\mu$m) and comb jellies (phylum Ctenophora, $1$-$15$\,cm).  Most diatoms move only by drifting, and are usually smaller than the Kolmogorov length scale of oceanic turbulence. Comb jellies locomote by a combination of active swimming and passive drifting, and their sizes typically lie within the inertial subrange of turbulence. Some diatoms can alter their nutrient uptake, settling velocity, or collision frequency by forming colonies with a variety of shapes.
In comb jellies, different body plans are correlated with different propulsion and predation modes, and we hypothesize that shape is very important for drifting-mode locomotion as well. A beautiful diversity of shapes has been observed for diatoms, diatom colonies, and comb jellies, extending from oblate to prolate forms.  Fore-aft symmetry is especially common in diatoms, and frequently found in comb jellies if we consider them in silhouette, \emph{i.e.} neglecting the placement of feeding appendages.  Axisymmetry is also common among diatoms and many orders of comb jellies.  Herein we focus on the basic question of passive shape-rotation interactions, upon which studies of active locomotion can build.

In this paper we investigate the effect of shape upon 
the rotation of particles in turbulence, their tumbling, and their spinning. 
We report results of direct numerical simulations (DNS) of particles rotating in turbulent flows, statistical-model calculations, and experiments.
These results enable us to characterize
how the orientational dynamics of rods and disks are qualitatively very different (Section \ref{sec:traj}).
But we also show that, despite these significant differences, the variance 
of angular velocity is  almost shape-independent (Section \ref{sec:ensemble}).
We show that these results are due to the inherent nature of the turbulence, 
by comparing the DNS results to those obtained in a random-flow model with finite
correlation length and time  (\Secref{randomflow}). Finally we discuss laboratory 
results quantifying the rotation rate of large particles in turbulence
(particles whose sizes fall in the inertial range of turbulence). These results
show  that the angular-velocity variance of large particles at relatively low aspect ratio is shape-independent too, 
thus extending our results concerning the dissipative range (\Secref{experiment}).

Before describing these results in detail we 
introduce the notation and briefly mention previous work in \Secref{background}.
Methods are discussed in \Secref{methods}.
\section{Background}\seclab{background}

We consider cylinders and spheroids. Both shapes are characterized by an axis of symmetry (length $2c$) and two other axes of equal lengths $2a\!=\!2b$. The aspect ratio is defined as $\ar=c/a$. 
The particle orientation is defined by the unit vector $\ve n$  that points along the axis of symmetry of the particle. The vector $\ve n$ evolves according to
\begin{equation}
\dot{\ve n} = \ve \omega \wedge \ve n\,.
\label{eq:tumble_rate}
\end{equation}
where $\ve \omega$ is the angular velocity of the particle and the dot denotes a time derivative.

The particle angular velocity $\ve \omega$ can be decomposed into components parallel and 
orthogonal to $\ve n$. The magnitude of the parallel component, $|\ve n\cdot\ve\omega|$, describes the rate at which the particle spins around its symmetry axis, 
the \lq spinning rate\rq{}. The magnitude of the orthogonal component, $|\dot{\ve n}|$, is called the 
\lq tumbling rate\rq{}. It is the combined rotation rate about the equatorial particle axes.

We consider the case of isotropic turbulence in the absence of external body forces. Thus the steady--state statistics of 
the angular velocity vector $\ve \omega$ is                                     isotropic. In particular  $\langle \ve \omega \rangle=0$. 
The variance of $\ve \omega$ is thus simply $\langle | \ve \omega|^2\rangle \equiv \langle \ve \omega \cdot\ve \omega\rangle$. 
Furthermore, since we consider particles with fore-aft symmetry,
the averages of $\ve n, \ve \omega \cdot \ve n$, and $\ve \omega \wedge \ve n$ must also vanish
because $-\ve n$ represents the same physical configuration as $\ve n$.
Thus the variances of the tumbling and spinning rates are simply $\langle |\dot{\ve n}|^2 \rangle$ and $\langle (\ve n \cdot \ve \omega)^2\rangle$.

From the above definitions it follows that the squared rotation, tumbling and spinning rates obey a kinematic relationship: the total rotation rate squared is the sum of the squared tumbling and spinning rates,
\begin{equation}
\eqnlab{decompose}
 | \ve \omega|^2  =  | \dot{\ve n}|^2+ | \ve n\cdot\ve \omega|^2\,.
\end{equation}
It follows by averaging \Eqnref{decompose} that the same relationship holds for the variances.

The dynamics of $\ve n$  have been studied by a number of authors. It has been shown that this orientation vector 
preferentially aligns with the strain and vorticity directions of the flow, depending on particle shape \cite{Par12,Pum11,Gus14,Ni14,Che13}. The tumbling rate is readily observed in experiments \cite{Par12,Ein13}, and its probability distribution and dependence on particle 
shape are known \cite{Par12,Gus14,Ni14,Che13}. Less is known about the spinning rate and the total angular velocity of axisymmetric particles suspended in turbulent flows. Our goals here are to qualitatively describe the dynamics and compare the distributions of these different rotation variables over a range of particle aspect ratios.  

\section{Methods} \seclab{methods}
\subsection{Direct numerical simulations}\seclab{dns}
Direct numerical simulations (DNS) of particle motion can be performed with one-way coupling of the fluid to the particle for the special case of infinitesimally--small neutrally--buoyant particles.
In this case, the particle center--of--mass is simply advected, and $\dot{\ve n}$ is described by equation (\ref{eq:tumble_rate}) using Jeffery's approximation \cite{Jef22} for the angular velocity:
\begin{align}
\dot {\ve x} &= \ve u(\ve x_t,t)\,,\label{eq:com}\\
\ve \omega &= \ve \Omega(\ve x_t, t) + \Lambda \ve n \wedge \ma S(\ve x_t, t)\ve n\label{eq:dotn}\,.
\end{align}
Here $\ve u(\ve x_t,t)$ is the fluid velocity at the particle position at time $t$. The vector $\ve \Omega$ equals half the vorticity and $\ma S = (\ma A + \ma A\transpose)/2$ is the strain-rate matrix, 
the symmetric part of the fluid--velocity--gradient matrix $\ma A$ with elements 
$A_{ij}=\partial u_i/\partial x_j$.
The antisymmetric part of $\ma A$ is denoted by $\ma O$, and $\ma O\ve n = \ve \Omega \wedge \ve n$.
The parameter  $\Lambda=(\ar^2-1)/(\ar^2+1)$ characterizes the shape of the particle:  $\Lambda=0$ for spheres,   $\Lambda=1$  for infinitely thin rods, and  $\Lambda=-1$ 
for infinitely thin disks. Most numerical studies that use this approach focus on rod-like particles \cite{Pum11,Ni14}, 
but some studies also consider oblate spheroids~\citep{Par12,Che13,Gus14}.
Eq.~(\ref{eq:dotn})  shows that spherical particles ($\Lambda=0$) respond only to the vorticity of the fluid, while non--spherical particles are affected by the fluid strain as well.  

The results shown below were obtained using time--series for $\ve u$ and $\ma A$ downloaded from the Johns Hopkins University turbulence database \cite{Li08,Yu12}. The database contains a DNS of forced, isotropic turbulence on a $1024\times 1024\times 1024$ grid at a Taylor--microscale Reynolds number of $\mbox{Re}_\lambda=433$. The particles are initialized at 
randomly chosen positions $\ve x$  and  orientations $\ve n$. Given  $\ve u(\ve x_t,t)$ 
and $\ma A(\ve x_t,t)$, particle position and orientation are updated according to Eqs. (\ref{eq:tumble_rate},\ref{eq:dotn},\ref{eq:com}) for approximately $45$ Kolmogorov times $\tau_{\rm K}$.  
We disregard the initial transient by discarding data corresponding to the first $10$\,$\tau_{\rm K}$ for each trajectory.
Distributions of the particle angular velocity and its spinning and tumbling rates are computed from the remaining data.  
The DNS shows that the variances of rotation, spinning, and tumbling rates rapidly approach their steady-state values, in most cases within $10$\,$\tau_{\rm K}$ (not shown).
But the results summarized in Appendix \secref{orthogonality} show that differences in the rotation between prolate and oblate particles may take
longer to develop for near--spherical particles. 

\subsection{Statistical-model calculations}
It is instructive to compare the DNS results for particle rotation in turbulent flows with those obtained for particles rotating in an isotropic homogeneous 
Gaussian random velocity field with appropriate correlation length and correlation time. This comparison shows which aspects of particle spinning and tumbling are influenced by the 
nature of turbulence, 
 and which aspects can be explained by a simple statistical model. An important difference between turbulence and a random Gaussian velocity field is that turbulence breaks time-reversal invariance, such that the fluid-velocity gradient matrix   $\ma A$ 
and its transpose $\ma A\transpose$ appear with different probabilities \cite{Che99}. 
In the statistical model, by contrast, $\ma A$ and $\ma A\transpose$  
appear with equal probabilities. This is important for our question because the orientational dynamics of rods and disks are determined
by $\ma A=\ma S+\ma O$ and $-\ma A\transpose=-\ma S+\ma O$, respectively. 
This follows from two observations. First 
the orientational equation of motion (\ref{eq:tumble_rate},\ref{eq:dotn}) 
can be recast as $\dot{\ve n} = (\ma O+\Lambda \ma S) \ve n- \Lambda (\ve n \cdot \ma S\ve n)\ve n$. Second, the non-linear term on right--hand side 
of this equation 
determines only the normalization of $\ve n$ but not its orientation (see Appendix \secref{orthogonality}).

Further differences between turbulence and random flow arise from the fact that turbulent flows exhibit much more violent vorticity fluctuations than random flows exhibit. This is important for our question because long-lived vortex 
structures \cite{She90} cause the particles to align, affecting the relation between spinning and tumbling rates 
\cite{Par12,Pum11,Gus14,Ni14,Che13}.

Our statistical model has two dimensionless parameters,
the shape factor $\Lambda$ and a second parameter that is formed out of the correlation length  $\eta$
of the fluid velocity, its correlation time $\tau$, and its typical speed $u_0$. 
This parameter is referred to as the Kubo number $\ku=u_0 \tau/\eta$. It is a dimensionless measure of the correlation time of the fluid velocity field. The incompressible random velocity field is 
represented as follows\cite{Gus14b}. 
We write $\ve u = \ve \nabla \wedge \ve A$, 
where $\ve A(\ve x,t)$ is a Gaussian random vector 
potential with zero mean, Gaussian spatial correlation function
with correlation length $\eta$, and exponential time correlations
$\langle A_i(x,t) A_j(x,0)\rangle = \delta_{ij} \exp(-t/\tau)/6$. 
This stochastic model is difficult to solve in closed form
because the orientational dynamics 
are determined by the Lagrangian correlations of the fluid-velocity gradients. 
The tumbling rate can be computed approximately by
perturbation theory \cite{Gus14,Gus14b}.
This gives rise to an expansion in the Kubo number. 
Up to the sixth order in $\ku$, the tumbling rate is given by\cite{Gus14}:
\begin{eqnarray}
\label{eq:p1}
&&\langle |\dot{\ve n}|^2\rangle \tau^2 = \ku^2(5+3\Lambda^2)/6+\ku^4\Lambda^2(5+3\Lambda^2)/4\\
&&+\ku^6 \Lambda^2(\!-25\!+\!4668\Lambda\!+\!45\Lambda^2\!+\!7236\Lambda^3\!+\!2484\Lambda^4)/864\,.\nn
\end{eqnarray}
The lowest-order term was obtained earlier \cite{Par12}. 
In a similar way the variance of the particle angular velocity can be computed. We find:
\begin{eqnarray}
&&\langle |{\ve \omega}|^2\rangle \tau^2=\ku^2(5+2\Lambda^2)/4+\ku^4\Lambda^2(5+3\Lambda^2)/4\nn\\
&&+\ku^6 \Lambda^2 (527+5\Lambda+804\Lambda^2+276\Lambda^4)/96
\end{eqnarray}
and
\begin{eqnarray}
\label{eq:p3}
\langle ({\ve n\cdot\ve \omega})^2\rangle \tau^2
&=&\langle ({\ve n\cdot\ve \Omega})^2\rangle \tau^2/4\\
&=&5\ku^2/12+25\ku^6\Lambda(1+3\Lambda)/864\,.\nn
\end{eqnarray}
The terms in these perturbation expansions contain only even powers of $\ku$. 
The perturbation series 
(\ref{eq:p1}) to (\ref{eq:p3})  are asymptotically divergent, \emph{i.e.} they diverge for any fixed value of 
$\ku$ but every partial sum of the series approaches the correct result as $\ku\to0$. To obtain accurate results at larger Kubo numbers requires resummation of the series. We have obtained the series expansions to order eight, 
making it possible to resum the series using Pad\'e{}-Borel resummation\cite{Gus13b}.

\subsection{Laboratory measurements}
The methods described above assume that particles are infinitesimally small, so that their rotation is always determined by the local fluid-velocity gradients. The rotation
of larger particles is influenced by their interactions with the non-linear fluid-velocity 
field.  Furthermore, the inertia of finite-size particles
feeds back to the fluid phase.  For an axisymmetric particle in simple shear flow, particle and fluid inertia make substantial contributions to the orientational motion \cite{Sub05,Sub06,Ein14}. But it is an open question to which extent weakly--inertial particles in turbulence behave similarly to inertia-less ones, or not.
The effect of particle inertia upon the orientational dynamics of particles in turbulence has
been calculated numerically \cite{Mar10} and for random flows analytically\cite{Gus14}, neglecting the effect of fluid inertia.  Two--way--coupled simulations that take into account fluid and particle inertia have been conducted for infinitesimally small particles \cite{zha13,and12}, but computing the effects of inertia becomes even more difficult for finite--sized particles.  Here we present, therefore, experimentally--measured rotation rates of large particles in turbulence, with sizes in the inertial sub-range of ambient turbulence.
\begin{table}
\begin{tabular}{ccccc}
\hline\hline
Height & Diameter              & $\ar = H/D$ & Volume  & Surface Area\\
 $2c$\,[mm] & $2a\!=\!2b$\,[mm]&                & [cm$^3$]& [cm$^2$] \\
\hline
4.77 $\pm$ 0.11 & 10.60 $\pm$ 0.13 & 0.45 $\pm$ 0.01 & 0.421 & 2.27 \\
8.24 $\pm$ 0.18 & 8.72 $\pm$ 0.06 & 0.95 $\pm$ 0.03 & 0.492 & 1.91\\
12.99 $\pm$ 0.14 & 6.41 $\pm$ 0.11 & 2.03 $\pm$ 0.04 & 0.420 & 1.48\\
18.91 $\pm$ 0.06 & 4.70 $\pm$ 0.04 & 4.03 $\pm$ 0.03 & 0.328 & 1.24\\
\hline
\hline
\end{tabular}
\caption{\label{tab:1} Dimensions of hydrogel cylinders; volume varies no more than $20\%$ 
around a mean of $0.415$ cm$^3$. Surface area varies no more than $30\%$ around a mean of $1.72\,$cm$^2$. Errors marked are standard error.
 }
\end{table}

Homogeneous isotropic turbulence is created in a $3$\, m$^3$ water tank using two facing arrays of randomly firing jets \cite{Bel14,Bel13}.  Turbulence in the test section has $\mbox{Re}_\lambda=310$, turbulent kinetic energy $6.5$ cm$^2$s$^{-2}$, integral length scale $8$cm, Kolmogorov length scale $\eta_{\rm K} =0.5$mm, and Kolmogorov time scale $\tau_K = 0.13$ s.  These scales were computed from Eulerian two--point velocity statistics computed with particle image velocimetry (PIV), specifically the autocovariance and second--order structure functions \cite{Bel13}.

Particles are added to the turbulent flow at a volume fraction of $0.1\%$, for which particle--particle collisions are negligible \cite{Elg94}.  Four types of particles are measured, with dimensions given in Table \ref{tab:1}.  Particles are $1\%$ denser than the ambient fluid, but stay in suspension due to the strength of the ambient turbulence.  Their quiescent--flow settling velocity normalized by the turbulent velocity scale is between 0.46 and 0.72, depending on the particle type.  

The particle Reynolds number in turbulent flow has several possible definitions \cite{Bell12b}; here we use the instantaneous slip velocity vector, computed by subtracting the particle center--of--mass velocity from the fluid velocity averaged over a 2D annulus surrounding the particle, exclusive of the immediate particle boundary layer.  The RMS magnitude of this slip velocity is close to 1 cm/s, across all particle types and regardless of the parameters used to define the outer bound of the fluid--averaging annulus.  From this, we compute a turbulent particle Reynolds number between 50 and 200, depending on the length scale that is used (the same range as the Reynolds number based on the quiescent settling velocity).  In the slip velocity, there is a small bias towards gravitational settling, with a mean value close to 2 mm/s for all particles.  
None of our measurements suggest that the particles are governed by Stokesian dynamics, but we can still compute a Stokesian response time as a point of reference. Using the radius of a sphere with equivalent volume to the cylinders (0.46 cm), the Stokesian time scale is $4.8$ s.  Computing the inertial--range time scale corresponding to a length of 0.46 cm gives a characteristic time scale $\tau_c = 0.57$ s.  Comparing these two time scales suggests that particles respond slowly compared to turbulent fluctuations at their length scale.

Particles are fabricated from hydrogel, in this case $0.4\%$ agarose by volume\cite{Byr13}.
Because hydrogel is clear and refractive--index--matched to water, we can use PIV to track the motion of tracers (hollow glass spheres) embedded within the particles.  
These tracers reveal the particle's rigid--body motion.  Applying stereoscopic PIV to the embedded tracers gives 3D velocity vectors on a 2D grid covering a planar slice through a particle.  From these data we compute the particle angular velocity using the equation for solid--body rotation: 
$\ve u_m-\ve u_n = \ve \omega \wedge (\ve x_m-\ve x_n)$ where $\ve u_m$ and $\ve u_n$ are velocity vectors at points $m$ and $n$ inside the particle, whose locations are $\ve x_m$ and $\ve x_n$.
This equation is solved using more than two vectors, to take advantage of all the data present within a particle's 
internal vector field.  Specifically, we use an optimization scheme based on vector triplets\cite{Bel12}.

Velocity vectors are calculated with multi-pass particle image velocimetry (PIV) based on cross-correlation of two-dimensional image subsets, followed by stereoscopic reconstruction of $\ve u(\ve x)$ within the measurement plane. The computations (and the supporting calibrations) are performed using the DaVis $7$ software package (Lavision Inc; Goettingen, Germany).  Tracers are illuminated by a $1$mm--thick laser light sheet (Quantel/Big Sky Lasers, $532$nm) passing through the flow and the particles.  Two cameras (Imager PRO-X, $1600\times 1200$ pixels, both fitted with a $105$mm Nikkor lens and a Scheimpflug/tilt adapter) focus on a subset of the light sheet of dimensions of 
$74.4$mm $\times$ $35.4$mm.  The cameras view the test section in a stereoscopic configuration through $35^\circ$ water-filled prisms mounted on the side of the tank to minimize distortion through the air-glass-water transition \cite{Bel14}.

\section{Small rods and disks in turbulence}
\seclab{smallparticles}

\begin{figure*}
\includegraphics{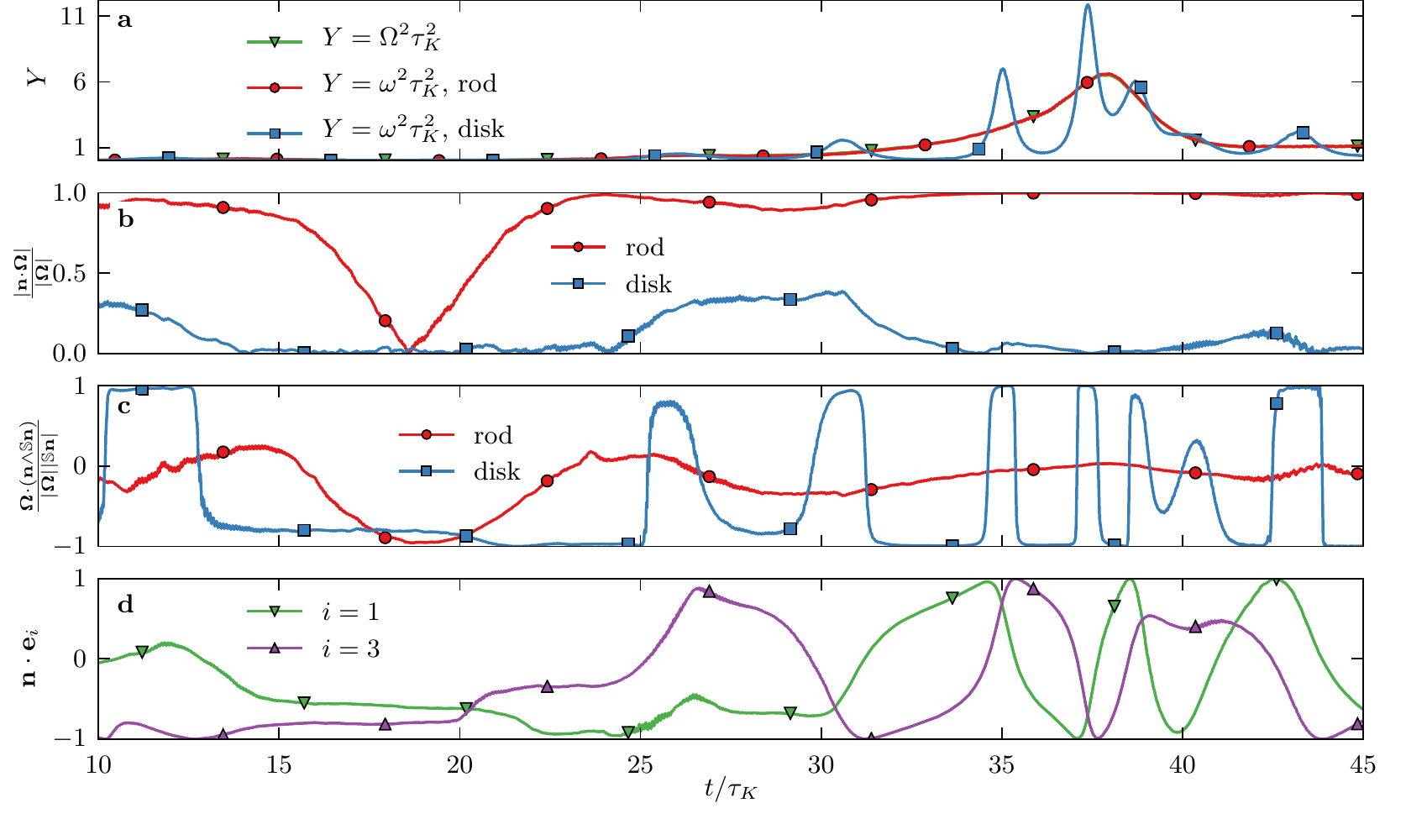}
\caption{\figlab{trajeks_35}DNS results for the instantaneous alignments and rotation rates for a disk and a rod as a function of time.
{\bf a} $|\ve \Omega|^2$ (green {\small$\color{figgreen}\blacktriangledown$}) and $|\ve \omega|^2$ 
as a function of time for disks (blue {\tiny$\color{figblue}\blacksquare$}) and rods (red $\color{figred} \bullet$).
{\bf b}  alignment of $\ve n$ with ${\ve \Omega}$ as a function of time for disks (blue {\tiny$\color{figblue}\blacksquare$}) and rods (red $\color{figred} \bullet$),
{\bf c} alignment of $\ve n \wedge \ma S\ve n$ with ${\ve \Omega}$ as a function of time for disks (blue {\tiny$\color{figblue}\blacksquare$}) and rods (red $\color{figred} \bullet$),
{\bf d} alignment of $\ve n_\textrm{disk}$ with $\ve e_1$ (green {\small$\color{figgreen}\blacktriangledown$}) and $\ve e_3$ (purple {\small$\color{figpurple}\blacktriangle$}) as a function of time.}
\end{figure*}

In this section we use DNS results for the orientational trajectories of
individual axisymmetric particles to characterize
the orientational dynamics of rods and disks in turbulence. This section
is divided into two parts. We first describe the orientational motion of individual
particles (Section \ref{sec:traj}) and then steady--state 
ensemble averages over many particle paths (Section \ref{sec:ensemble}).

\subsection{Orientational trajectories of individual particles}
\label{sec:traj}
In \Figref{trajeks_35} we show a typical trajectory which brings the tracer particle through a region of intense vorticity (starting at $t\!=\!\unit[30]{\tau_K}$).
A tracer disk ($\Lambda\!=\!-1$) and a tracer rod ($\Lambda \!=\! 1$) both follow the same center-of-mass trajectory, but their orientational dynamics are  very different. The blue squares ({\tiny$\color{figblue}\blacksquare$}) represent the disk, and the red circles ($\color{figred} \bullet$) the rod.

\Figref{trajeks_35}\textbf{a} shows that for this trajectory the magnitude of the angular velocity of the rod is almost identical to $|\ve \Omega|$, half the flow vorticity (green {\small$\color{figgreen}\blacktriangledown$}). 
The angular velocity of the disk, by contrast, fluctuates strongly around the fluid angular velocity. The fluctuations occur on a time scale comparable to the Kolmogorov time. This qualitative difference 
between the orientational motion of  rods and disks in regions of strong vorticity 
is explained by the preferential alignment of particles to the fluid-velocity gradients.

DNS shows that rods (or material lines) and vorticity in turbulence tend 
to align with each other. This is usually attributed to the fact that the respective equations of motion are closely related
\cite{Dre1992,Pum11,Ni14}.  
This behavior is borne out by the trajectory in \Figref{trajeks_35}\textbf{b}. 
It follows from the alignment of rods with vorticity that the spinning rate of rods $(\ve n \cdot \ve \omega)^2$ is approximately equal to the fluid rotation rate $|\ve \Omega|^2$. 
In addition, a rod tumbles, staying closely aligned with $\ve \Omega$. But this tumbling motion is slow on the \lq vorticity time scale\rq ~$\sim\!\unit[10]{\tau_K}$.

The orientational dynamics of the disk ($\Lambda=-1$) are expected to be very different from those of a rod ($\Lambda=1$) 
because $\ve n_{\rm disk}$ is driven by $-\ma S+\ma O$ while $\ve n_{\rm rod}$ is driven by $\ma S+ \ma O$.  \Figref{trajeks_35}\textbf{b} shows that
the disk aligns so that its symmetry axis is in the plane orthogonal to that of the rod. This fact
is a simple consequence of the form of the equation of motion (\ref{eq:tumble_rate},\ref{eq:dotn}) and the observation that the Lyapunov exponent of incompressible turbulent flow is
positive. This argument is explained in detail in Appendix \secref{orthogonality}. It is consistent with the intuition that the symmetry vector of a disk lies in the plane orthogonal
to that of a rod because a long axis of both particles is being aligned by the Lagrangian fluid stretching\cite{Mar14}.

Because the vector $\ndisk$ lies in the plane perpendicular to $\nrod$, it follows that $\ndisk$ tends to be found perpendicular to the vorticity direction. This behavior can be seen in \Figref{trajeks_35}\textbf{b} as mentioned above.  This alignment leads to a vorticity-induced tumbling rate $\ve \Omega \wedge \ndisk$ and a correspondingly weak spinning rate. But for disks also the contribution from the strain, $\ve n \wedge \ma S \ve n$, is strong. It alternates between enhancing and opposing the rotation of $\ndisk$ around the vorticity direction. This follows from the curious observation that material lines tend to instantaneously align with the second eigen--direction $\ve e_2$ of the strain-rate matrix $\ma S$ (that is, with the smaller of the two positive eigenvalues) in regions of high vorticity\cite{Ash1987,Pum11}. Since $\ve n_{\rm disk}$ tends to be perpendicular to $\ve n_{\rm rod}$ we expect that $\ndisk$ (and therefore also $\ma S\ndisk$) tends to lie in the plane spanned by $\ve e_1$ and $\ve e_3$, the two strongest eigen--directions of $\ma S$. The resulting cross product is parallel to $\ve e_2$ and significant in magnitude. This alignment between $\ve \Omega$ and $\ndisk\wedge\ma S\ve \ndisk$, shown in \Figref{trajeks_35}\textbf{c}, is responsible for the fluctuations in total angular velocity of the disk.

\Figref{trajeks_35}\textbf{d} illustrates that the symmetry vector of a disk tumbles in the plane spanned by $\ve e_1$ and $\ve e_3$ in regions of strong vorticity.  The two largest rotation events for the disks at $t=35\tauK$ and $t=37\tauK$ both coincide with events where $\ve n$ becomes aligned with $\ve e_1$ and then rapidly rotates to become aligned with $\ve e_3$. Comparing \Figref{trajeks_35}\textbf{a} and \Figref{trajeks_35}\textbf{d} shows that when the vorticity is weak, there is little or no instantaneous alignment of $\ve n$ with the eigen-system of the strain-rate matrix\cite{Ni14} (see also \Figref{trajeks_1}\textbf{d} and \Figref{trajeks_12}\textbf{d} in Appendix~\secref{orthogonality}).

\subsection{Ensemble averages}
\label{sec:ensemble}

\begin{figure}
\includegraphics{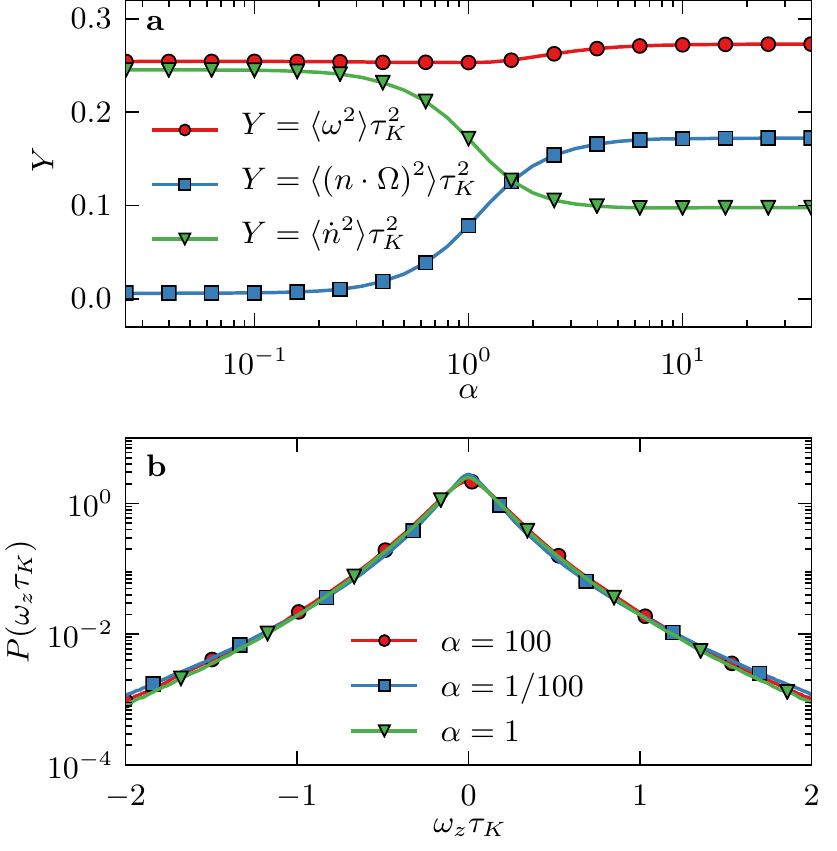}
\caption{\figlab{DNS}{\bf a} DNS results for the variances of the angular velocity (red $\color{figred} \bullet$), and the spinning (blue {\tiny$\color{figblue}\blacksquare$}) and tumbling rates (green {\small$\color{figgreen}\blacktriangledown$}) as a function of the particle aspect ratio
$\ar$, computed for tracer spheroids in homogeneous isotropic turbulence. The data for the tumbling rate is similar to the data shown in Fig. 3 in Ref. \citenum{Par12} and Fig. 2 in Ref. \citenum{Gus14}.
The rates are made dimensionless by rescaling with the Kolmogorov time. 
{\bf b}
Distribution of the $z$-component of the angular velocity from DNS, for different aspect ratios: rod, $\ar=100$ (red $\color{figred} \bullet$); disk, 
$\ar=1/100$ (blue {\tiny$\color{figblue}\blacksquare$}); spheres $\ar=1$ (green {\small$\color{figgreen}\blacktriangledown$}). 
The rates are made dimensionless by rescaling with the Kolmogorov time.
}
\end{figure}

In the previous section we showed that the angular velocities of disks and rods are very different, due to the differing alignments of their symmetry axes to the vorticity vector. Nevertheless, upon averaging over many trajectories we find that the average rotation rate $\langle |\ve \omega|^2\rangle$ is almost independent of shape. In \Figref{DNS} we show the average total rotation, spinning and tumbling rates as a function of aspect ratio $\ar$. The curves are obtained using the DNS results described above.   Disks rotate on average just like spheres, while rods rotate slightly faster.   However, the tumbling and spinning rates show strong dependence on shape.  This indicates, from \Eqnref{decompose}, that the tumbling rate and its complement, the spinning rate, must vary as a function of  $\ar$ in such a manner that their sum is nearly $\ar$-independent. In other words, rods spin more than they tumble, while disks tumble more than they spin.  The tumbling and spinning rates are especially divergent for disks whose spinning rate is almost zero. The effect of shape on tumbling and spinning is strongest for small departures
from $\ar=1$.  Outside of the range $0.1<\ar<10$, particle rotations saturate at constant values, remaining insensitive to further changes in aspect ratio.

\Figref{DNS}  shows that the violent fluctuations of the rotation rate of disks around the fluid rotation rate (visible in \Figref{trajeks_35}\textbf{a}) average to zero, 
while the much smaller fluctuations of the rotation rate of rods average 
to a small positive contribution. The nature of these fluctuations depends crucially on the precise dynamics of rods, disks and vorticity in relation to the local strain eigen--system \cite{Dre1992}.
Revealing the combined dynamics of these vectors is an important goal in the study of Lagrangian turbulence and the dynamics of suspended particles. We believe that the study of disks may add to the picture of Lagrangian turbulence. 

\section{Small particles in random flows}\seclab{randomflow}
\begin{figure}
\includegraphics{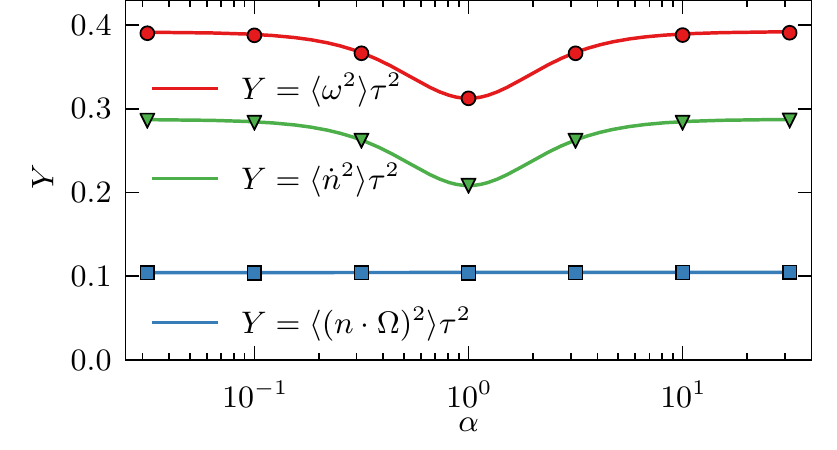}
\caption{\figlab{randomflow}
Variances of the angular velocity (red $\color{figred} \bullet$), of the  tumbling rate (green {\small$\color{figgreen}\blacktriangledown$}), and of the spinning rate (blue {\tiny$\color{figblue}\blacksquare$}) as a function of the particle aspect ratio $\ar$
for the statistical model, computed by resumming the perturbation expansions (\ref{eq:p1}) to (\ref{eq:p3}), lines. Markers show results of numerical simulations of the statistical model. Parameter: $\ku=0.5$. }
\end{figure}

We argue above that the observed rotation rates non-trivially arise from the 
specific properties of the turbulent velocity gradient tensor, as observed in a Lagrangian frame.
In this section, we demonstrate this claim by analyzing the orientational dynamics in a random-flow model. The random flow we present has finite space and time correlations, corresponding to the Kolmogorov scales in turbulence. Solving the random flow model enables us to answer the question of which observations are \emph{not} due to turbulence, but simply features of the equations of motion\cite{Gus14b}.

Tumbling, spinning, and rotation variances for the statistical model are given in Eqs. (\ref{eq:p1}) to (\ref{eq:p3}). We have also computed the $\ku^8$-contribution to these expressions (not shown). 
This allows us to resum the perturbation series using Pad\'e{}-Borel resummation\cite{Gus13b}.
The resulting rotation, spinning and tumbling variances are shown in \Figref{randomflow}  as a function of particle shape. In comparison to the turbulence result, we find three notable differences. First, the average total rotation rate is shape-dependent: nearly spherical particles rotate less than non-spherical particles. Second, the shape-dependence is almost completely encoded in the tumbling rate, while the spinning rate is almost independent of particle shape. This means that preferential alignment exists but is weak in a random flow. This in turn is because vorticity and strain are not directly related, as is the case in turbulence, but only through preferential sampling along trajectories. Third, to a very good approximation rods and disk rotate, spin and tumble alike. This is a consequence of the fact that the statistical model is time-reversal invariant and does not allow for long-lived vortex structures \cite{Gus14}. Finally we note that the symmetry axes of co-located disks and rods do become perpendicular to each other also in the random flow (see Appendix~\secref{orthogonality}); however, none of them significantly aligns with the vorticity direction.

\section{Large particles in turbulence}\seclab{experiment}

\begin{figure}
\includegraphics{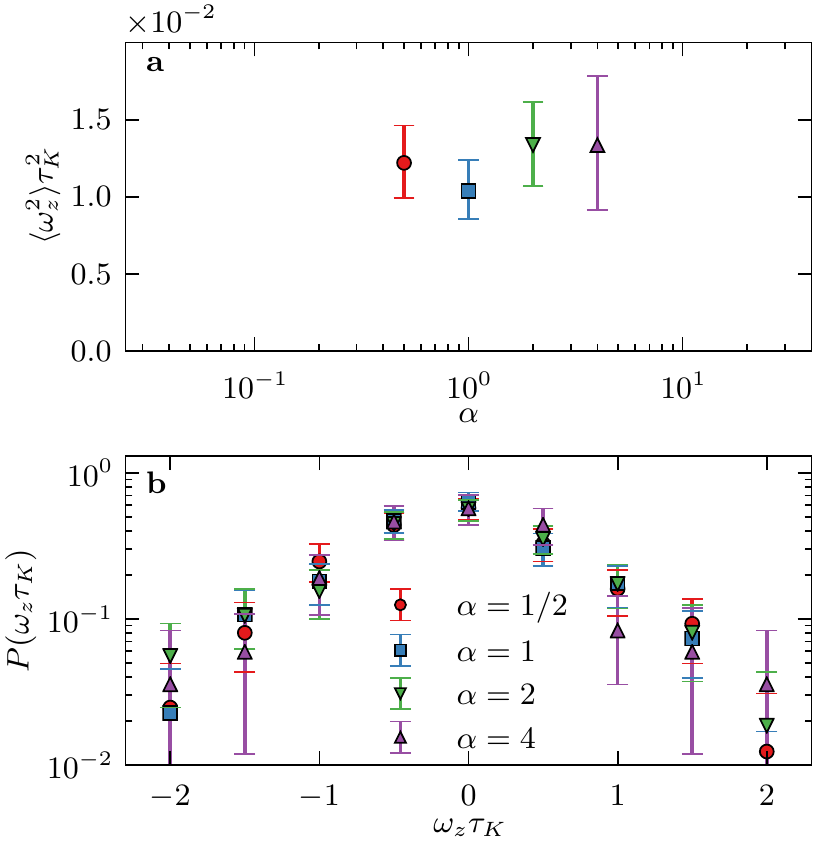}
\caption{\figlab{exp}{\bf a} Variance of the $z$-component of the particle angular velocity (inverse seconds squared) as a function of the cylinder aspect ratio $\ar$,
measured for large cylinders in laboratory turbulence.  $95\%$
confidence intervals are computed via bootstrap. {\bf b} Distribution of $z$-component of the particle angular velocity for the four different cylinder aspect ratios given in Table 1; $150$-$400$ samples are used in each of the four curves. Aspect ratio $\ar = 1/2$ (red $\color{figred} \bullet$), $1$ (blue {\tiny$\color{figblue}\blacksquare$}), $2$ (green {\small$\color{figgreen}\blacktriangledown$}), $4$ (purple {\small$\color{figpurple}\blacktriangle$}).}
\end{figure}

In the preceding sections we considered tracer particles in turbulence and random flows. In this section we address the question of larger particles by a laboratory experiment.

Compared to tracer particles, larger particles are affected by several additional mechanisms. First, they have particle inertia in both translational and orientational degrees of freedom. Second, the forces and torques upon larger particles include the effects of fluid inertia. Third, the finite-sized particles sample 
the non-linear velocity field on the scale of the particle size. All of these mechanisms potentially invalidate the arguments concerning preferential alignment, so we do not know, \emph{a priori}, whether large particles will show shape--dependent rotation that resembles our observations for small particles.

Our results for
the total rotation rate of large particles are shown in \Figref{exp}. Within experimental error, the shape-dependence of the average total rotation rate is consistent with our findings in \Secref{smallparticles}. This remarkable fact remains to be explained.
It is expected that the rotation-rate variance is smaller for large particles (because they average over fluid-velocity gradients). This is borne out by the experiment: it shows a variance that is approximately a factor of $6$ smaller than the DNS variance (note that we plot $\omega_z^2$ for the experiment and $|\ve \omega|^2$ for the DNS, which differ by a factor of $3$).  
Previous work\cite{Par14} has
successfully predicted the tumbling rate variance of high-aspect-ratio rods in the inertial subrange,   showing that rod length controls the tumbling rate.  However, extending this approach to low-aspect-ratio rods and disks is non-trivial. If rod rotation scaling were based on one lengthscale only, we would expect a significant difference in the angular velocity variance for the particles with $\ar=2$ and $\ar=4$, which here have $2c$=16$\eta$ and $2c$=38$\eta$ respectively.  However, we observe no significant change in angular velocity variance. This suggests that dynamics of inertial, low-aspect-ratio rods are not accurately predicted by a single lengthscale.

\section{Conclusions}
In this paper we have analyzed the rotation of axisymmetric particles in turbulence by experiments, direct numerical simulations, and random-flow model calculations. We have found that
disks rotate very differently from rods in regions of intense vorticity in turbulence.
While the symmetry axis of a rod follows closely the second strain eigenvector $\ve e_2$ and vorticity, the symmetry axis of the disk tumbles in the plane spanned by $\ve e_1$ and $\ve e_3$, the strain eigenvectors corresponding to the largest extensional and largest compressional eigenvalues.  Rods spin around their own symmetry axis at a rate of half the vorticity, while fluid vorticity and strain act to make disks tumble.  In other words, because of their different alignment with respect to fluid vorticity, rods tend to spin more than they tumble, while disks tend to tumble more than they spin.

This has important implications for the instantaneous 
rotation dynamics in turbulence. The strain makes only a small contribution to the
rotation of rods, while for disks it makes
a large contribution: it tends to rotate the symmetry vector of disks around the vorticity vector, 
sometimes decelerating, sometimes accelerating the rotation.

Despite this qualitative difference, the variance of the total rotation rate is almost independent of shape. In fact, the rotation rate variance of a disk is neither more nor less than that of a sphere. 
This exact equivalence is unexpected. Also, prolate
particles rotate with nearly the same angular velocity  variance as disks and spheres. This
is borne out by both DNS and experiments, for small and large particles respectively.

We also demonstrated by a random-flow calculation that our observations in turbulence are not a feature of the equations of motion, but depend on the distinguishing statistical features of turbulence in a Lagrangian frame. In the random-flow model the rotation rate variance depends on shape, and the preferential alignment between fluid vorticity and the symmetry axes of particles is very weak.

Several aspects of particle rotation in turbulence remain to be understood.
An important open question concerns the alignment of disks, rods, and vorticity with the eigen--system $\ve e_j(t)$ of the strain-rate matrix, $j=1,2,3$. The time-delayed correlations $\langle \ve \Omega(t)\cdot\ve e_1(0)\rangle$ shed some light on the vorticity dynamics\cite{Xu2011}, but is there a corresponding `pirouette effect' for disks?
A second important question is to understand the effect of finite particle sizes. Our results from DNS and
the statistical model pertain to tracer particles. Finite-size corrections and the effects of particle and fluid inertia to the particle-rotation rate remain to be understood.
A third question to consider is the implication of our results concerning the rotation and alignment of less symmetric particles such as ellipsoids with three distinct moments of inertia or non-ellipsoidal particles. 

A major motivation for this investigation is the effect of shape upon the dynamics of planktonic organisms.  The relationships between the shape of a planktonic organism, its kinematics in flow, and its biological success are quite complex, but our results can contribute a few facts to this ongoing investigation.  First, shape does not control how much angular velocity an organism inherits from the ambient turbulence on average, 
but it does control how this angular velocity is distributed about the organism's principal axes.  This may impact swimming behavior according to the directionality of the propulsive system employed (directional, as in copepods, or omnidirectional, as in cydippid ctenophores). Second, it is tempting to label planktonic species as either \lq spinners\rq~or \lq tumblers\rq~based on their shape, but this nomenclature cannot be taken too literally: our results indicate that shape can only \emph{emphasize} spinning or tumbling, it cannot select one exclusively.  In other words, even the most extreme tumblers show a fair bit of spinning, and vice versa.  Potential reasons for an organism to emphasize tumbling over spinning may include mass transfer, swimming, or gyrotaxis \cite{del14,hil97}.  Third, if an organism or colony were to emphasize spinning or tumbling by changing its shape, the greatest marginal return occurs when aspect ratio is near unity. Once an organism departs significantly (by a factor of $10$) 
from this ratio, further shape changes have no effect on rotation.  
Spheroidal body plans with aspect ratios near unity ($1 < \alpha < 2$) are commonly observed in cydippid ctenophores, which places them in the range for which  small changes in shape will greatly influence rotation.  This is suggestive that shape may play a major role in their behavior and locomotion.  Of course, continued cross--disciplinary study is needed to elucidate the full impact of body shape on plankton biology in complex flows. Herein we have described some of the passive physical mechanisms involved, and future work may build upon this foundation, investigating how and whether aquatic organisms take advantage of these mechanisms.

{\em Acknowledgements}. We thank NORDITA in Stockholm for hospitality during the workshop Particles in Flows - Fundamentals and Applications where this manuscript was begun.  This material is based upon work supported by the National Science Foundation under Grants No. DGE-1106400 (MB), OCE-1334788 (EV), and  DMR-1208990 (GV) and by the Vetenskapsr\aa{}det, and the G\"o{}ran Gustafsson Foundation for Research in the Natural Sciences and Medicine (JE,KG,BM).

\appendix
\section{Relation between the alignment of disks and rods}\seclab{orthogonality}
\begin{figure}
\label{fig:q}
\includegraphics{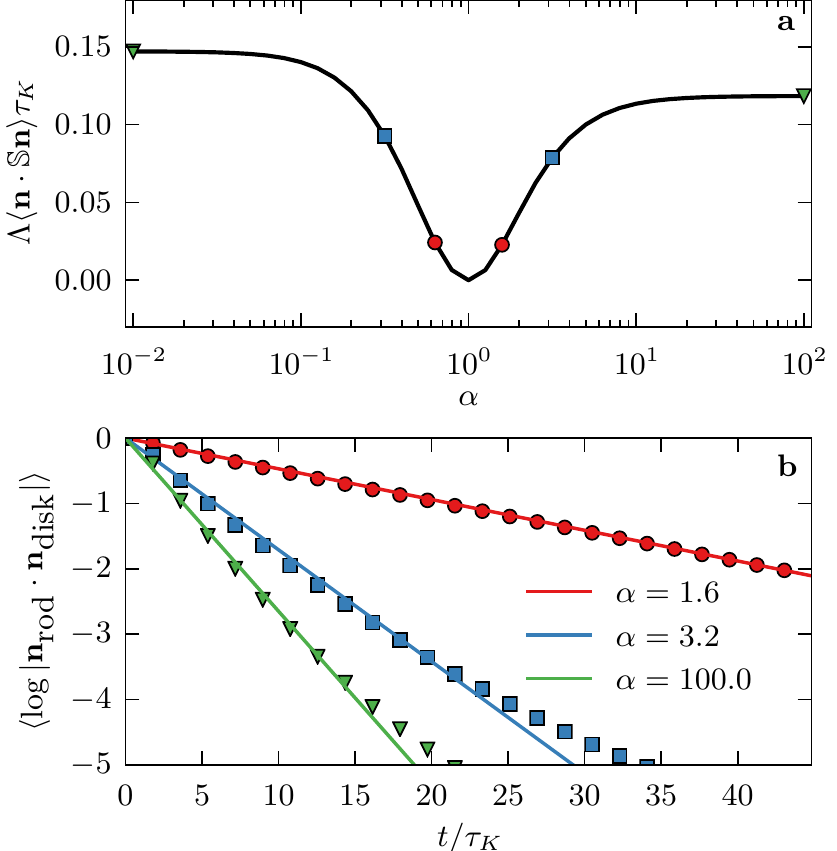}
\caption{{\bf a} DNS results for $\Lambda \langle \ve n\cdot \ma S\ve n \rangle$ as a function of
the particle-shape parameter $\ar$, solid line. The symbols indicate the values used to compute
the solid lines in the lower panel of this figure.
{\bf b} Alignment of $\nrod$ and $\ndisk$, which tends to zero as time increases. At $t=0$, $\nrod$ and $\ndisk$ are parallel along $\hat{\ve z}$.
DNS results (symbols) compared
to expectation (exponential growth rates taken from panel {\bf a}, lines). Parameters
$\ar = 100$ and $1/100$ ({\small$\color{figgreen}\blacktriangledown$}), $\ar = 3.2$ and $1/3.2$ ({\tiny$\color{figblue}\blacksquare$}), and $\ar = 1.6$ and $1/1.6$
($\color{figred} \bullet$).}
\end{figure}
Consider the time evolution of the vector $\nrod$ for a rod with shape factor $\Lambda$ and the corresponding vector $\ndisk$ for a disk with shape factor $-\Lambda$.
Assume that the center--of--mass of the rod and the disk take the same paths
through the turbulent flow. How do the symmetry vectors of the rod and the disk
align with respect to each other?  Recall the equations of motion
(\ref{eq:tumble_rate}) and (\ref{eq:dotn}). They can be rewritten as
\begin{align}
        \dot{\ve n} &= \ve \Omega \wedge \ve n + \Lambda
\ma S \ve n - \Lambda  (\ve n\cdot \ma S \ve n) \ve n\,.
\end{align}
It follows that the cosine of the angle between the two vectors $\ve n_{\rm rod}$ and $\ve n_{\rm disk}$ for rods and disks
with shape factors $\Lambda$ and $-\Lambda$ evolves
according to
\begin{align}
        &\frac{\rd}{\rd t}\left[\nrod\cdot\ndisk\right] =
        \dot{\ve n}_\mathrm{rod}\cdot\ndisk +
        \nrod\cdot \dot{\ve n}_\mathrm{disk} \\
        &= -\Lambda \nrod\cdot \ndisk (\nrod \cdot\ma S \nrod - \ndisk\cdot \ma S \ndisk)
\nn
\end{align}
provided that the centers of mass of both particles follow the same path through the fluid.
Put differently,
\begin{align}
\eqnlab{lyap}
        \frac{\rd}{\rd t}\log|\nrod\cdot \ndisk| &= \\
&\hspace*{-1cm}- (\Lambda \nrod \cdot \ma S \nrod - \Lambda \ndisk\cdot \ma S \ndisk)\,.\nonumber
\end{align}
The right--hand side of this equation, evaluated along particle trajectories,
tends to be negative because of the way the symmetry vectors of rods and
disks align with the eigensystem of the strain $\ma S$.  This is why we cannot discuss $\ve n$ and $\ma O$ only.
The steady--state average of the right--hand side is negative; DNS
results confirm this (Fig. 5{\bf a}).
It follows that the angle
between $\ve n_{\rm rod}$ and $\ve n_{\rm disk}$  must decrease
as a function of time. This is demonstrated by the DNS results
shown in Fig. 5{\bf b}.

We note that the quantity $\Lambda\langle \ve n \cdot \ma S \ve n \rangle$
(shown in Fig. 5{\bf a}) is the exponential growth rate
(Lyapunov exponent) of a vector $\ve q$ evolving according to
\begin{align}\eqnlab{dotq}
        \dot{\ve q} &= (\ma O + \Lambda \ma S) \ve q, \qquad\ve n = \ve q/|\ve q|.
\end{align}
The vector $\ve q$ points in the same direction as $\ve n$ but it is not normalised
\cite{Bre62,Wil09}, $\ve n = \ve q/|{\ve q}|$. In order to compute the orientational dynamics it is sufficient to
consider the vector $\ve q$
that obeys the linear equation of motion \eqnref{dotq}.
In the limit of $\alpha \rightarrow \infty$ (rods or material lines, $\Lambda\to 1$)
this Lyapunov exponent was computed in earlier direct numerical simulations \cite{Girimaji1990}
These simulations at smaller Reynolds numbers (${\rm Re}_\lambda = 38,63$, and $90$) obtained a
Lyapunov exponent of $0.13$ in units of $\tau_{\rm K}$ which is in fairly good agreement with our data in Fig. 5{\bf a} for large aspect ratio.

If the Lyapunov exponents of $\ve q_\pm$ with shape factors $\pm \Lambda$ sum to a positive number for a given flow, the symmetry vectors of rods and disks
must become orthogonal to each other. As Fig. 5{\bf a} shows,
this is the case for isotropic turbulence and this is consistent with
the arguments summarized in the previous paragraph.

It is also the case for the random-flow model
discussed in \Secref{randomflow}, where to $O(\ku^4)$
\begin{equation}
\Lambda
 \langle \ve n\cdot \ma S\ve n\rangle  \tau = 2 \ku^2 \Lambda^2  - \ku^4 \Lambda^2(9\Lambda^2 + 16)/3+\ldots\,.
\end{equation}

\section{Supplementary Figures}
\begin{figure}[p]
\begin{overpic}{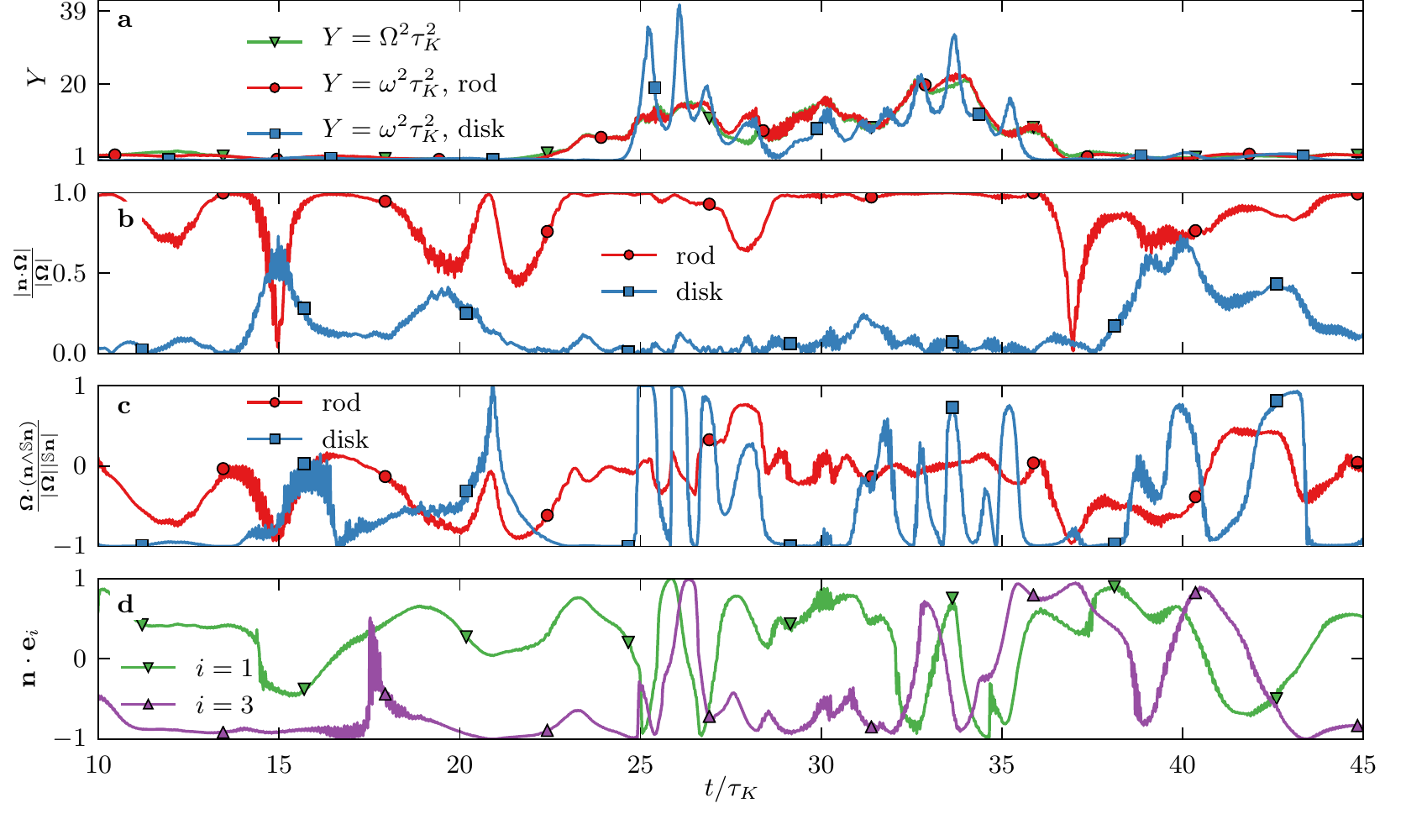}
\end{overpic}
\caption{\label{fig:trajeks_1} DNS results for the instantaneous alignments and rotation rates of a disk and a rod as a function of time.
{\bf a} Fluid angular velocity variance $|\ve \Omega|^2$ (green {\small$\color{figgreen}\blacktriangledown$}) and particle angular velocity variance $|\ve \omega|^2$
as a function of time for disk (blue {\tiny$\color{figblue}\blacksquare$}) and rod (red $\color{figred} \bullet$).
{\bf b}  alignment of $\ve n$ with ${\ve \Omega}$ as a function of time for disk (blue {\tiny$\color{figblue}\blacksquare$}) and rod (red $\color{figred} \bullet$),
{\bf c} alignment of $\ve n \wedge \ma S\ve n$ with ${\ve \Omega}$ as a function of time for disk (blue {\tiny$\color{figblue}\blacksquare$}) and rod (red $\color{figred} \bullet$),
{\bf d} alignment of $\ve n_\textrm{disk}$ with $\ve e_1$ (green {\small$\color{figgreen}\blacktriangledown$}) and $\ve e_3$ (purple {\small$\color{figpurple}\blacktriangle$}) as a function of time.}
\end{figure}

\begin{figure}[p]
\begin{overpic}{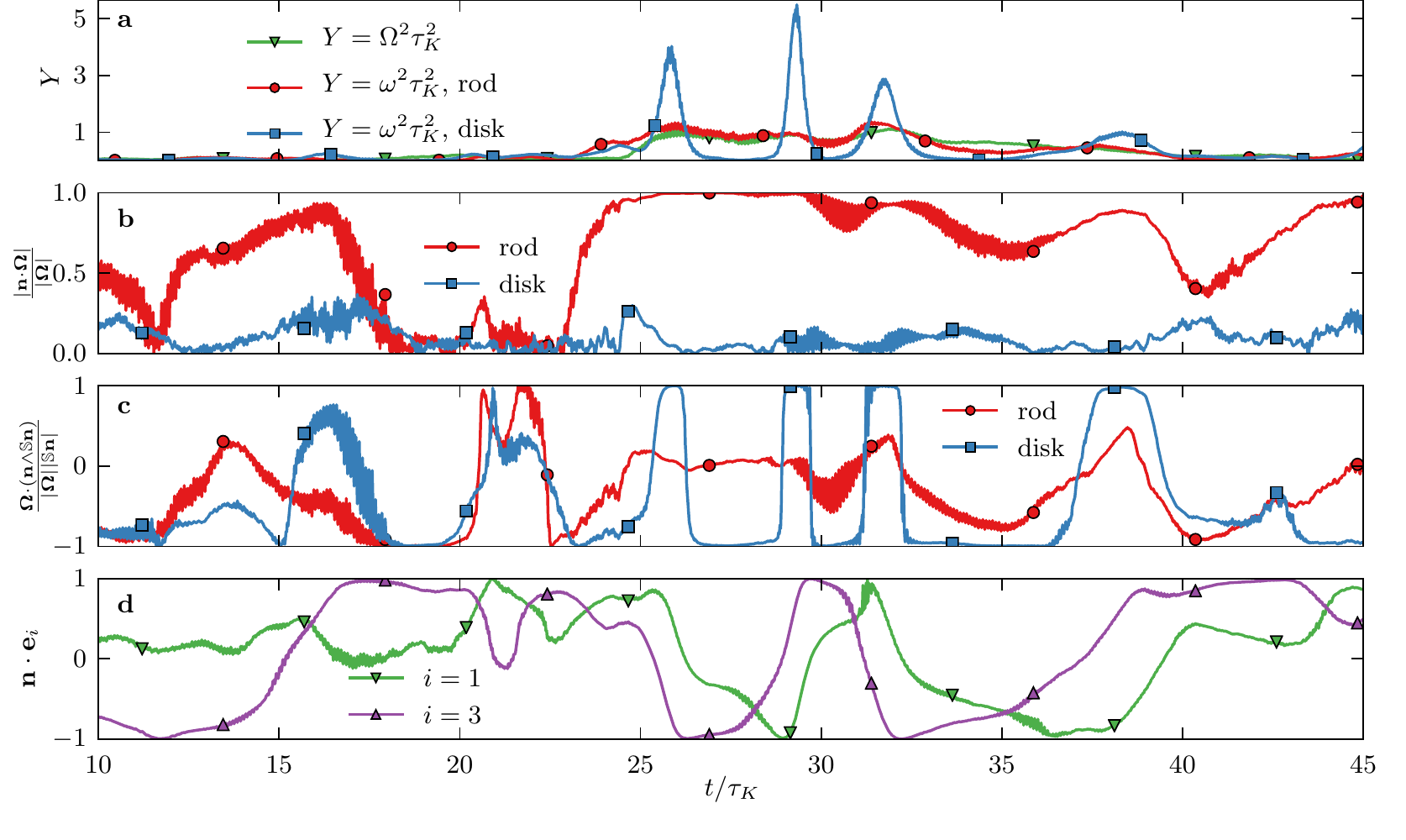}
\end{overpic}
\caption{\label{fig:trajeks_12} DNS results for the instantaneous alignments and rotation rates for a disk and a rod as a function of time.
{\bf a} $|\ve \Omega|^2$ (green {\small$\color{figgreen}\blacktriangledown$}) and $|\ve \omega|^2$
as a function of time for disks (blue {\tiny$\color{figblue}\blacksquare$}) and rods (red $\color{figred} \bullet$).
{\bf b}  alignment of $\ve n$ with ${\ve \Omega}$ as a function of time for disks (blue {\tiny$\color{figblue}\blacksquare$}) and rods (red $\color{figred} \bullet$),
{\bf c} alignment of $\ve n \wedge \ma S\ve n$ with ${\ve \Omega}$ as a function of time for disks (blue {\tiny$\color{figblue}\blacksquare$}) and rods (red $\color{figred} \bullet$),
{\bf d} alignment of $\ve n_\textrm{disk}$ with $\ve e_1$ (green {\small$\color{figgreen}\blacktriangledown$}) and $\ve e_3$ (purple {\small$\color{figpurple}\blacktriangle$}) as a function of time.}
\end{figure}

This appendix contains further examples of orientational trajectories
of disks and rods in turbulence obtained from the DNS described in the main text.
 The rapid oscillations seen in Figs. \ref{fig:trajeks_1} and \ref{fig:trajeks_12}
are due to artifacts in numerically determining the fluid-velocity gradients.
That such oscillations are weaker
in Fig. \ref{fig:trajeks_35} is a coincidence.

\end{document}